\newfam\scrfam
\batchmode\font\tenscr=rsfs10 \errorstopmode
\ifx\tenscr\nullfont
        \message{rsfs script font not available. Replacing with calligraphic.}
        \def\scr{\cal}

\else   
        \font\sevenscr=rsfs7
        \font\fivescr=rsfs5
        \skewchar\tenscr='177 \skewchar\sevenscr='177 \skewchar\fivescr='177
        \textfont\scrfam=\tenscr \scriptfont\scrfam=\sevenscr
        \scriptscriptfont\scrfam=\fivescr
        \def\scr{\fam\scrfam}
        \def\cal{\scr}
\fi
\newfam\msbfam
\batchmode\font\twelvemsb=msbm10 scaled\magstep1 \errorstopmode
\ifx\twelvemsb\nullfont\def\Bbb{\bf}
        \message{Blackboard bold not available. Replacing with boldface.}
\else   \catcode`\@=11
        \font\tenmsb=msbm10 \font\sevenmsb=msbm7 \font\fivemsb=msbm5
        \textfont\msbfam=\tenmsb
        \scriptfont\msbfam=\sevenmsb \scriptscriptfont\msbfam=\fivemsb
        \def\Bbb{\relax\expandafter\Bbb@}
        \def\Bbb@#1{{\Bbb@@{#1}}}
        \def\Bbb@@#1{\fam\msbfam\relax#1}
        \catcode`\@=\active
\fi
        \font\eightrm=cmr8              \def\xrm{\eightrm}
        \font\eightbf=cmbx8             \def\xbf{\eightbf}
    	\font\eightit=cmti10 at 8pt     \def\xit{\eightit}
        \font\eighttt=cmtt8             \def\xtt{\eighttt}
        \font\eightcp=cmcsc8
        \font\eighti=cmmi8              \def\xold{\eighti}
        \font\teni=cmmi10               \def\old{\teni}
        
        \font\tentt=cmtt10
        \font\twelverm=cmr12
        \font\twelvecp=cmcsc10 scaled\magstep1
        \font\fourteencp=cmcsc10 scaled\magstep2

        \font\eightmath=cmmi8
	\font\eightsym=cmsy8

\def\ss{\scriptstyle}

\headline={\ifnum\pageno=1\hfill\else
{\eightcp M. Cederwall, B.E.W. Nilsson, P. Sundell:
        ``An action for the super-5-brane$\ldots$''}
                \dotfill{ }{\old\folio}\fi}
\def\makeheadline{\vbox to 0pt{\vss\noindent\the\headline\break
\hbox to\hsize{\hfill}}
        \vskip2\baselineskip}
\newcount\foottest
\foottest=0
\def\makefootline{
        \ifnum\foottest>0
                \ifnum\foottest=1
                        \footline={\footlineone}
                \fi
                \ifnum\foottest=2
                        \footline={\footlineone\footlinetwo}
                \fi
                \baselineskip=.8cm\vtop{\the\footline}
                \global\foottest=0
        \fi
        }
\def\footnote#1#2{${}^#1$\hskip-3pt
	\ifnum\foottest=1
        \def\footlinetwo{\hfill\break
        \vtop{\baselineskip=9pt
        \indent ${}^#1$ \vtop{\hsize=14cm\noindent\xrm #2}}}\foottest=2
        \fi
        \ifnum\foottest=0
        \def\footlineone{\vtop{\baselineskip=9pt
        \hrule width.6\hsize\hfill\break
        \indent ${}^#1$ \vtop{\hsize=14cm\noindent\xrm #2}}
        \vskip-.7\baselineskip}
        \foottest=1
        \fi
	}
\newcount\refcount
\refcount=0
\newwrite\refwrite
\def\ref#1#2{\global\advance\refcount by 1
        \xdef#1{{\old\the\refcount}}
	\ifnum\the\refcount=1
        	\immediate\openout\refwrite=\jobname.refs
        \fi
        \immediate\write\refwrite
                {\item{[{\xold\the\refcount}]} #2\hfill\par\vskip-2pt}}
\def\xref#1#2#3{\global\advance\refcount by 1
        \xdef#1{{\old\the\refcount}}
	\xdef#2{{\xold\the\refcount}}
	\ifnum\the\refcount=1
        	\immediate\openout\refwrite=\jobname.refs
        \fi
        \immediate\write\refwrite
                {\item{[{\xold\the\refcount}]} #3\hfill\par\vskip-2pt}}
\def\refout{\catcode`\@=11
        \xrm\immediate\closeout\refwrite
        \vskip2\baselineskip
        {\noindent\twelvecp References}\hfill\vskip\baselineskip
        \baselineskip=.75\baselineskip
        \input\jobname.refs
        \baselineskip=4\baselineskip \divide\baselineskip by 3
        \catcode`\@=\active\rm}
\newcount\eqcount
\eqcount=0
\def\Eqn#1{\global\advance\eqcount by 1
        \xdef#1{{\old\the\eqcount}}
        \eqno({\oldstyle\the\eqcount})
        }
\def\eqn{\global\advance\eqcount by 1
        \eqno({\oldstyle\the\eqcount})
        }
\def\multi{\vskip-.5\belowdisplayskip \global\advance\eqcount by 1}
\def\intermulti{\vskip-\belowdisplayskip}
\def\multieq#1#2{\xdef#1{{\old\the\eqcount#2}}
        \eqno{({\oldstyle\the\eqcount#2})}}
\parskip=3.5pt plus .3pt minus .3pt
\baselineskip=14pt plus .1pt minus .05pt
\lineskip=.5pt plus .05pt minus .05pt
\lineskiplimit=.5pt
\abovedisplayskip=18pt plus 4pt minus 2pt
\belowdisplayskip=\abovedisplayskip
\hsize=15cm
\vsize=19cm
\hoffset=1cm
\voffset=1.8cm
\frenchspacing
\def\/{\over}
\def\*{\partial}
\def\a{\alpha}
\def\b{\beta}

\def\d{\delta}
\def\e{\varepsilon}

\def\g{\gamma}
\def\k{\kappa}
\def\l{\lambda}

\def\w{\!\wedge\!}

\def\F{{\cal F}}

\def\C{{\Bbb C}}

\def\punkt{\,.}
\def\komma{\,,}
\def\.{.\hskip-1pt }

\def\+{\!+\!}
\def\={\!=\!}
\def\minus{\!-\!}
\def\>{\!>\!}

\def\half{{\lower2pt\hbox{\eightrm 1}\/\raise2pt\hbox{\eightrm 2}}}
\def\frac#1{{\lower2pt\hbox{\eightrm1}\/\raise2pt\hbox{\eightrm#1}}}
\def\tfrac#1{{\lower2pt\hbox{\trm1}\/\raise2pt\hbox{\trm#1}}}
\def\genfrac#1#2{{\lower2pt\hbox{\eightrm#1}\/\raise2pt\hbox{\eightrm#2}}}

\def\tr{\hbox{\rm tr\,}}

\def\II{\hbox{I\hskip-0.6pt I}}

\def\C{{\cal C}}
\def\A{{\cal A}}
\def\F{{\cal F}}

\def\H{{\cal H}}

\def\L{{\cal L}}

\def\={\!=\!}
\def\half{{\lower2.5pt\hbox{\eightrm 1}\/\raise2.5pt\hbox{\eightrm 2}}}
\def\fraction#1{{\lower2.5pt\hbox{\eightrm 1}\/\raise2.5pt\hbox{\eightrm #1}}}
\def\Fraction#1#2{
        {\lower2.5pt\hbox{\eightrm #1}\/\raise2.5pt\hbox{\eightrm #2}}}
\def\ihalf{{\lower2.5pt\hbox{\eightmath i}\/\raise2.7pt\hbox{\eightrm 2}}}
\def\ifrac#1{{\lower2.5pt\hbox{\eightmath i}\/\raise2.5pt\hbox{\eightrm#1}}}
\def\tr{\hbox{\rm tr}}

\def\WZ{Wess--Zumino}
\def\CS{Chern--Simons}
\def\BI{Bianchi identity}

%
%
%

\null\vskip-1cm
\hbox to\hsize{\hfill G\"oteborg-ITP-97-16}
\hbox to\hsize{\hfill CTP-TAMU-51/97}
\hbox to\hsize{\hfill\tt hep-th/9712059}
\hbox to\hsize{\hfill December, 1997}
\hbox to\hsize{\hfill revised March, 1998}

\vskip2cm
\centerline{\fourteencp An action for the super-5-brane in D=11 supergravity}
\vskip4pt
\vskip\parskip
\centerline{\twelvecp}

\vfill
\centerline{\twelverm Martin Cederwall, Bengt E.W. Nilsson}

\vskip.5cm
\centerline{\it Institute for Theoretical Physics}
\centerline{\it G\"oteborg University and Chalmers University of Technology }
\centerline{\it S-412 96 G\"oteborg, Sweden}

\vskip.5cm
\catcode`\@=11
\centerline{\tentt tfemc,tfebn@fy.chalmers.se}
\catcode`\@=\active

\vskip1.2cm
\centerline{\twelverm Per Sundell}

\vskip.5cm
\centerline{\it Center for Theoretical Physics}
\centerline{\it Texas A \&\ M University}
\centerline{\it College Station, TX 77843, USA}

\vskip.5cm
\catcode`\@=11
\centerline{\tentt per@chaos.tamu.edu}
\catcode`\@=\active

\vfill

{\narrower\noindent\underbar{\it Abstract:}
An alternative path is taken for deriving an action for the 
supersymmetric 5-brane in 11 dimensions. Selfduality does not follow
from the action, but is consistent with the equations of motion for arbitrary
supergravity backgrounds. 
The action involves a 2-form as well as a 5-form 
world-volume potential; inclusion
of the latter makes the action, as well as the non-linear selfduality
relation for the 3-form field strength, polynomial. 
The requirement of invariance under $\kappa$-transformations
determines the form of the selfduality relation, as well as the action.
The formulation is shown to be equivalent to earlier formulations
of 5-brane dynamics.
\smallskip}

\vfill
\eject

\def\nl{\hfill\break\indent}
\def\nlni{\hfill\break}

\ref\ElevenSG{E. Cremmer, B. Julia and  J. Scherk,
	{\xit ``Supergravity theory in eleven-dimensions''},
	\nl Phys.Lett. {\xbf76B} ({\xold1978}) {\xold409};\nlni
	L.~Brink and P.~Howe, {\xit ``Eleven-dimensional supergravity 
	on the mass-shell in superspace''},
	\nl Phys.~Lett.~{\xbf 91B} ({\xold1980}) {\xold384};\nlni
	E.~Cremmer and S.~Ferrara, 
	{\xit ``Formulation of eleven-dimensional supergravity 
	in superspace''},\nl Phys.~Lett.~{\xbf 91B} ({\xold1980}) {\xold61}.}
\ref\Witten{E.~Witten, 
	{\xit ``String theory dynamics in various dimensions''},
	Nucl.~Phys.~{\xbf B443}
	({\xold1995}) {\xold85} [{\xtt hep-th/9503124}].}
\ref\Duality{C.M. Hull and P.K. Townsend,
        {\xit ``Unity of superstring dualities''},
        Nucl. Phys. {\xbf B438} ({\xold1995}) {\xold109}
        [{\xtt hep-th/9410167}];\nlni
        J.H. Schwarz, {\xit ``The power of M theory''},
        Phys. Lett. {\xbf B367} ({\xold1996}) {\xold97}
        [{\xtt hep-th/9510086}];\nlni
        A.~Sen, {\xit ``Unification of string dualities''},
	{\xtt hep-th/9609176};\nlni
	P.K. Townsend, {\xit ``Four lectures on M-theory''},
        {\xtt hep-th/9612121}.}
\ref\Guven{R.~G\"uven, 
	{\xit ``Black p-brane solutions of D=11 supergravity theory''}, 
	Phys.~Lett.~{\xbf B276} ({\xold1992}) {\xold49}.}
\ref\Supermembrane{E.~Bergshoeff, E.~Sezgin and P.K.~Townsend, 
	\nl {\xit ``Supermembranes and eleven-dimensional supergravity''},
	Phys.~Lett.~{\xbf B189} ({\xold1987}) {\xold75};
	\nl {\xit ``Properties of the eleven-dimensional supermembrane 
	Theory''}, Ann.~Phys. {\xbf 185} ({\xold1988}) {\xold330}.}
\ref\Mtheory{B. de Wit, J. Hoppe and H. Nicolai,
	{\xit ``On the quantum mechanics of supermembranes''},\nl
	Nucl. Phys. {\xbf B305} ({\xold1988}) {\xold545};\nlni 
	P.K. Townsend, {\xit ``D-branes from M-branes''}, 
	Phys. Lett.~{\xbf 373B} ({\xold1996}) {\xold68} 
	[{\xtt hep-th/9512062}];\nlni
	T.~Banks, W.~Fischler, S.H.~Shenker and L.~Susskind,
	\nl{\xit ``M theory as a matrix model: a conjecture''},
	Phys. Rev. {\xbf D55} ({\xold1997}) {\xold5112} 
		[{\xtt hep-th/9610043}].}
\ref\SchwarzFivebrane{M. Aganagic, J. Park, C. Popescu and J.H. Schwarz,\nl
	{\xit ``World-volume action of the M theory five-brane''},
	{\xtt hep-th/9701166}.}
\ref\DualLagr{B. McClain, Y.S. Wu and F. Yu,\nl 
	{\xit ``Covariant quantization of chiral bosons 
		and OSp(1,1{\eightsym\char'152}2) symmetry''},
	Nucl. Phys. {\xbf B343} ({\xold1990}) {\xold689};\nlni
	I. Bengtsson and A. Kleppe, {\xit ``On chiral p-forms''},
	{\xtt hep-th/9609102};\nlni
	N. Berkovits, {\xit ``Manifest electromagnetic duality 
		in closed superstring field theory''},\nl
	Phys. Lett. {\xbf B388} ({\xold1996}) {\xold743} 
		[{\xtt hep-th/9607070}];\nlni
	P. Pasti, D. Sorokin and M. Tonin,
        {\xit ``On Lorentz invariant actions for chiral p-forms''},\nl
        Phys. Rev. {\xbf D55} ({\xold1997}) {\xold6292}
        [{\xtt hep-th/9611100}].}
\ref\FivebraneLagr{
	I. Bandos, K. Lechner, A. Nurmagambetov, P. Pasti, D. Sorokin
        and M. Tonin,\nl {\xit ``Covariant action for the super-five-brane
                of M-theory''},
        Phys. Rev. Lett. {\xbf78} ({\xold1997}) {\xold4332}
                [{\xtt hep-th/9701037}];\nlni
	I. Bandos, N. Berkovits and D. Sorokin,
	\nl{\xit ``Duality-symmetric eleven-dimensional 
		supergravity and its coupling to M-branes''},
	{\xtt hep-th/9711055}.}
\ref\HSWFivebrane{P.S.~Howe and E.~Sezgin, {\xit ``D=11, p=5''},
        Phys.~Lett.~{\xbf B394} ({\xold1997}) {\xold62}
        [{\xtt hep-th/9611008}];\nlni
        P.S.~Howe, E.~Sezgin and P.C.~West,\nl
        {\xit ``Covariant field equations of the M theory five-brane''},
        Phys.~Lett.~{\xbf B399} ({\xold1997}) {\xold49}
        [{\xtt hep-th/9702008}].}
\ref\ACGHN{T. Adawi, M. Cederwall, U. Gran, M. Holm and B.E.W. Nilsson,\nl
	{\xit ``Superembeddings, non-linear supersymmetry and 5-branes''},
	{\xtt hep-th/9711203}.}
\ref\WittenFivebrane{E. Witten, 
	{\xit ``Five-brane effective action in M-theory''},
	{\xtt hep-th/9610234}.}
\ref\WVStrings{P.S. Howe, N.D. Lambert and P.C. West,\nl
	{\xit ``The self-dual string soliton''},
	{\xtt hep-th/9709014};\nl
	{\xit ``The threebrane soliton of the M-fivebrane''},
	{\xtt hep-th/9710033}.}
\ref\Bonora{L. Bonora, C.S. Chu and M. Rinaldi,
	{\xit ``Perturbative anomalies of the M-5-brane''},
	{\xtt hep-th/9710063}.}
\ref\CW{M. Cederwall and A. Westerberg,\nl
        {\xit ``World-volume fields, SL(2;Z) and duality:
        the type IIB 3-brane''}, {\xtt hep-th/9710007}.}
\ref\BLT{P.K. Townsend, {\xit ``Worldsheet electromagnetism and the superstring
        tension''}, Phys. Lett. {\xbf 277B} (1992) 285;\nlni
        E. Bergshoeff, L.A.J. London and P.K. Townsend,
        {\xit ``Space-time scale-invariance and the super-p-brane''},\nl
        Class. Quantum Grav. {\xbf 9} (1992) 2545 [{\xtt hep-th/9206026}].}
\ref\PKT{P.K. Townsend, {\xit ``Membrane tension and manifest IIB S-duality''},
        {\xtt hep-th/9705160}.}
\xref\CT\xCT{M. Cederwall and P.K. Townsend,
        {\xit ``The manifestly Sl(2;Z)-covariant superstring''},\nl
        JHEP {\xbf09} ({\xold1997}) {\xold003} [{\xtt hep-th/9709002}].}
\xref\BRO\xBRO{E. Bergshoeff, M. de Roo and T. Ortin,      
        {\xit ``The eleven-dimensional five-brane''},\nl
        Phys.~Lett.~{\xbf B386} ({\xold1996}) {\xold85}
        [{\xtt hep-th/9606118}].}
\xref\SixTensor\xSixTensor{P.S.~Howe, E.~Sezgin and P.C.~West,      
        {\xit ``The six-dimensional self-dual tensor''},\nl
        Phys.~Lett.~{\xbf B400} ({\xold1997}) {\xold255}
        [{\xtt hep-th/9702111}].}


\noindent 11-dimensional supergravity [\ElevenSG] is believed 
to be the low-energy effective
theory for M-theory. Arising as a strong coupling limit of type \II A
superstring theory [\Witten], it is connected to the entire web of dualities
relating different vacua of string theory/M-theory 
(see e.g. refs. [\Witten,\Duality]). 
The extended objects in 11-dimensional supergravity seem to be closely
connected to non-perturbative properties of the theory. 
The ones carrying non-gravitational charges are a membrane and 
a 5-brane, and in addition there are the 
KK (or gravitational) branes, which are a wave and a KK-monopole (6-brane).
The field
configuration [\Guven] corresponding to the membrane [\Supermembrane], 
acting as an electric source for the 3-form potential, is singular,
and the membrane, though never properly quantised, may prove to constitute
the elementary excitations of M-theory [\Mtheory]. The 5-brane, on the
other hand, is a solitonic, non-singular, field configuration, carrying
magnetic charge with respect to the 3-form.
A world-volume theory for a 5-brane is thought of as a low-energy theory
for the moduli of such solitonic solutions.

Due to the fact that the world-volume theory contains chiral bosons,
namely a (non-linearly) selfdual field strength, there are difficulties
with a lagrangian formulation. There are a number of different routes for 
circumventing the problem. One is to give up manifest Lorentz 
invariance [\SchwarzFivebrane].
Another is to use one of the existing techniques
for lagrangian formulations of theories with chiral bosons [\DualLagr].
These techniques were applied to the 11-dimensional 5-brane 
in ref. [\FivebraneLagr]. The formulation contains
auxiliary fields, whose main raison d'{\^e}tre seems to be to parametrise
the ``non-covariance'' of ref. [\SchwarzFivebrane]. Although the methods
of refs. [\SchwarzFivebrane,\FivebraneLagr] work locally on the world-volume,
there may be global obstructions.
Another interesting way of deriving the equations of motion is to consider
geometric properties of the
embedding of the world-volume superspace into target superspace 
[\HSWFivebrane,\ACGHN].

A different approach, which will be followed in this paper,
was suggested by Witten [\WittenFivebrane]. One may
write down a lagrangian whose equations
of motion do not imply selfduality, but allow it. The allowed form will
be unique due to the couplings to background fields. One then needs
a prescription for the implementation of this selfduality in a path integral.
Ref. [\WittenFivebrane] discusses this issue in the linearised model,
and we will not have anything further to say about that point.
We will concentrate on giving the full non-linear version of 
the action in this approach,
and to show, by explicit calculation, that it indeed follows from demanding
the correct amount of supersymmetry on the world-volume.
The obtained action is quite simple, and might be useful in situations
where an effective 5-brane action is needed.
Although some of the aspects of 5-brane dynamics are well understood
and some may have to await a more profound understanding of M-theory,
there are remaining questions, e.g. concerning world-volume solitons
[\WVStrings] and anomaly cancellations [\WittenFivebrane,\Bonora],
that should have appropriate answers in the supergravity limit.

Some words on the notation adopted in this paper: the conventions 
used for forms on superspace are the ordinary ones,
where the exterior derivative acts by wedge product from the right.
Complex conjugation does not revert the order of fermions, which results
in the absence of factors of $i$, found in most of the literature.

The only antisymmetric 
tensor field in $D$=11 supergravity is the 3-form potential
$C$, with field strength $H\=dC$. As usual when treating dynamics of
``higher-dimensional'' branes, it is convenient to include dual
field strengths in the background. We therefore also have a 6-form
potential $\C$ with field strength $\H=d\C+{1\/2}C\w H$, so that its
\BI\ reads  
$$
d\H-\half H\w H=0\punkt\Eqn\HBianchi
$$
The second term
is due to consistency of duality with the presence of the \CS\ term
$\sim\int C\w H\w H$. Analogous consistency conditions will be crucial
for determining the form of the 5-brane action and its background coupling.
The 5-brane is formulated as a 6-dimensional world-volume embedded
in 11-dimensional superspace, so the above identities are taken as being
forms on superspace. When demonstrating $\k$-symmetry, we will need
to specify the  
gauge-invariant background quantities carrying at least one spinor index.
The constraints containing non-vanishing field components relevant
for the $\k$-variation are (with suitable normalisation)
$$
\eqalign{
&{T_{\a\b}}^a=2(\g^a)_{\a\b}\komma\cr
&H_{ab\a\b}=2(\g_{ab})_{\a\b}\komma\cr
&\H_{abcde\a\b}=2(\g_{abcde})_{\a\b}\komma\cr}\Eqn\THH
$$
where $T$ is the superspace torsion.

Following ref. [\CW], we now introduce one world-volume 
antisymmetric tensor field for
each target space one, generically coupled to the latter with the 
D-brane type coupling ``$F\=dA\minus C$''. In the present case, we thus have
a 2-form potential $A$ and a 5-form $\A$. When the target space fields
obey modified Bianchi identities like eq. (\HBianchi) for $\H$, so
will the world-volume ones (the corresponding case for type \II B
was constructed in ref. [\CW]). Demanding that the world-volume field strengths
are invariant under both world-volume and target space gauge transformations
determines their form in terms of the potentials (up to trivial rescalings)
as
$$
\eqalign{
F&=dA-C\komma\cr
\F&=d\A-\C-\half A\w H\komma\cr}\Eqn\WVFields
$$
so that the Bianchi identities read
$$
\eqalign{
dF&=-H\komma\cr
d\F&=-\H-\half F\w H\punkt\cr}\Eqn\WVBianchis
$$
It is a straightforward exercise to show that these fields indeed are
invariant under 
$$
\eqalign{
\d C&=dL\komma\cr
\d\C&=d\L-\half L\w H\komma\cr
\d A&=L+dl\komma\cr
\d\A&=\L+d\ell+\half l\w H\punkt\cr}\eqn
$$
where $L$ and $\L$ are target superspace 2-form and 5-form gauge parameters,
while $l$ and $\ell$ are world-volume 1- and 4-forms. We use identical
symbols for target superspace forms and their pullbacks (which of course
are bosonic forms on the world-volume).

When considering $\k$-transformations, these are induced by local
translations of the world-volume in fermionic target space directions as
$\d_\k Z^M\=\k^\a{E_\a}{}^M$,
which results in the transformation of the pullback of a target space
form $\d_\k\Omega=\L_\k\Omega\equiv(i_\k d+di_\k)\Omega$.
The parameter $\k$ will, as usual, be constrained to contain only
half a spinor worth of independent components.
The transformations of the world-volume fields $A$ and $\A$ must be specified
so that their field strengths transform gauge-invariantly. 
These transformations are
$$
\left.
\matrix{\d_\k A=i_\k C\komma\hfill\cr
\d_\k\A=i_\k\C+\half A\w i_\k H\komma\hfill\cr}
\right\}
\quad\Longrightarrow\quad
\left\{
\matrix{\d_\k F=-i_\k H\komma\hfill\cr
\d_\k\F=-i_\k\H-\half F\w i_\k H\punkt\hfill\cr}
\right.\eqn
$$

We want to use the field strengths $F$ and $\F$ to build an action for
the 5-brane. The r\^ole of $\F$ is twofold. It replaces the \WZ\ term,
which in traditional brane actions gives the electric coupling of a
$p$-brane to a ($p$+1)-form potential. It also renders the action 
polynomial, as will soon be demonstrated. The entire procedure described
in the following could be performed without the $\F$ field and with
a \WZ\ term. We find the present formalism simpler and more appealing.

Following the procedure of refs. [\BLT,\PKT,\CT,\CW], we make an ansatz for the
action of the form
$$
S=\int\!d^6\xi\,\sqrt{-g}\,\l\left(1+\Phi(F)-({*}\F)^2\right)\komma
\Eqn\Actionansatz
$$
where $\Phi$ is some yet undetermined function of $F$ (with the lower
indices of the $F$'s contracted by inverse metrics), whose form will
be determined from symmetry arguments. We might have included a numerical
constant in front of the last term; it would then be determined to 1
for our actual normalisation of the fields in eq. (\WVFields).
The field $\l$ is a scalar Lagrange multiplier\footnote\dagger{The factor 
$\ss\sqrt{-g}$ is inessential for the dynamics of the 5-brane, and can be 
omitted if one uses a scalar density Lagrange multiplier. In ref. [\xCT], a
specific choice for the $\ss g$-dependence was made, whose sole consequence
was that a tensionless string was included in the spectrum.}, 
whose equation of motion enforces
$$
{*}\F=\pm\sqrt{1+\Phi(F)}\punkt\Eqn\Fsquarevalue
$$
The two solutions correspond to selfduality
or anti-selfduality for $F$ and charge plus or minus 1, i.e., a 5-brane
or an anti-5-brane. 

We should remark that the inclusion of the 6-form 
field strength is by no means necessary --- one could as well work with
an action with a square root type kinetic term plus a \WZ\ term. In ref.
[\PKT], a parallel was drawn between the existence of world-volume forms
of maximal rank and the the possibility of branes to possess boundaries.
It was noted that the absence of boundaries for the 5-brane would seem
to provide an argument against the type of formulation given in this paper.
It would be interesting to
understand whether or not the presence of the 6-form has more profound
implications; at the moment we do not know, but simply use it because
it is calculationally convenient.

The equations of motion for the world-volume potentials are
\multi
$$
d(\l{*}K)-\l({*}\F)H=0\komma\multieq\AtwoEOM a
$$
\intermulti
$$
d(\l{*}\F)=0\komma\multieq\AfiveEOM b
$$
where we have introduced the field $K\={\*\Phi\/\*F}$, and where the
dependence of $\F$ on $A$ is responsible for the second term in the
equation of motion (\AtwoEOM) for $A$. 
Taking the constraint following from variation
of $\l$ into consideration, the equation of motion (\AfiveEOM) for $\A$ simply
determines $\l$ up to a constant factor. 
A crucial point is now to compare the equation of motion
for $A$ to its \BI, thereby determining what kind of selfduality is
consistent with the ansatz\footnote*{The corresponding problem, without
the introduction of a world-volume 5-form potential, was formulated
and solved to lowest order in {\eighti F}
in ref. [\xBRO].}. It follows immediately that consistency of
the identification of Bianchi identities and equations of motion demands
the selfduality condition to read 
$$
{*}K=-({*}\F)F\punkt\Eqn\KDuality
$$
$K$ will typically be given as $F\+(\hbox{higher terms})$, 
and one way to continue
would be to look for explicit expressions for $K$ that turn eq. (\KDuality)
into a self-consistent relation, i.e., one that expresses one of the
linear chirality components of $F$ in terms of the other one. We will
however take another direction --- by considering $\k$-symmetry, the
function $\Phi$ in the action, and thereby the selfduality relation,
will be determined. The so determined selfduality will be consistent.

It is amusing to note that although the action is not specified with respect
to the dependence on the field $F$, any variation (including variations
w.r.t. the metric) will contain $K$, which
by the selfduality relation (\KDuality) may be reexpressed in terms of $F$.
The plan is therefore to perform a $\k$-variation, express the variation
of the action entirely in $F$ and $\F$, and look for conditions on these
fields that makes the variation vanish.
We may actually restrict the variation to the factor 
$\Psi\equiv1\+\Phi(F)\minus({*}\F)^2$,
since variation of the factors in front of it will yield terms 
proportional to $\Psi$ that vanish due to the constraint $\Psi\!\approx\!0$
(so called 1.5 order formalism), or compensated by the appropriate 
transformation of $\l$.

Performing the $\k$-variation and using the known relation between the
number of $F$'s and $g$'s in $\Phi$ gives
$$
\eqalign{
\d\Psi
&=\frac6K^{ijk}\d F_{ijk}-\frac4{F^i}_{kl}K^{jkl}\d g_{ij}\cr
&+\genfrac2{6!}\F^{ijklmn}\d\F_{ijklmn}
-\frac{5!}\F^{iklmnp}{\F^j}_{klmnp}\d g_{ij}\punkt\cr}\eqn
$$
Inserting the background fields of
eq. (\THH) gives the explicit expression in terms of the spinor $\k$:
$$
\d_\k\Psi
=4({*}\F)\bar E_i\,\Bigl[\,
	\frac{5!$\ss\sqrt{-g}$}\e^{ijklmn}\g_{jklmn}
+\half{*}F^{ijk}\g_{jk}+\frac4{F^{(i}}_{kl}{*}F^{j)kl}\g_j
+({*}\F)\g^i\,\Bigr]\,\kappa\punkt\Eqn\KappaVariation
$$
The task is now to find a matrix $M^\a{}_\b$ of half rank, so that
(\KappaVariation) vanishes for $\k\=M\zeta$. An explicit calculation,
where eq. (\KappaVariation) is expanded in antisymmetric products of 
$\g$-matrices of different ranks, using the most general ansatz for $M$,
shows that one necessarily must have
$$
\k=\left[\,{*}\F-\frac{$\ss\sqrt{-g}$}\e^{ijklmn}
\left(\frac{6!}\g_{ijklmn}
-\frac{2(3!)$\ss{}^2$}F_{ijk}\g_{lmn}\right)\right]\,\zeta\Eqn\Proj
$$
for some spinor $\zeta$.
Two other necessary conditions come out of the calculation, namely
the form of the selfduality relation,
$$
-({*}\F){*}F_{ijk}=F_{ijk}-\half {q_{[i}}^lF_{jk]l}+
\frac{12}\tr kF_{ijk}\komma\Eqn\SD
$$
where\footnote\dagger{The matrix {\eighti k} looks formally similar to
the one denoted by the same symbol in ref. [\xSixTensor] (denoted {\eighti r}
below), but since
our {\eighti F} is non-linearly selfdual, it contains a trace in addition
to the traceless part.} $k_{ij}\={1\/2}{F_i}^{kl}F_{jkl}$, 
$q_{ij}\=k_{ij}-\fraction6g_{ij}\tr k$ , and
$$
({*}\F)^2=1+\frac{24}\tr q^2-\frac{144}(\tr k)^2\punkt\Eqn\Fisone
$$
Due to selfduality, the matrix $q$ fulfills $q^2={1\/6}\tr q^2$,
and the the antisymmetrisation made explicit in eq. (\SD) is automatic.
By comparing the form of the selfduality relation to the
condition (\KDuality), the function $\Phi$ in the action is determined:
$$
\Phi=\frac6\tr k-\frac{24}\tr q^2+\frac{144}(\tr k)^2\komma\Eqn\Phiis
$$
so that, by the equation of motion for the Lagrange multiplier,
$$
({*}\F)^2=\bigl(1+\frac{12}\tr k\bigr)^2-\frac{24}\tr q^2\punkt\Eqn\Fistwo
$$
Consistency of these two expressions for ${*}\F$ (eqs. (\Fisone) and 
(\Fistwo)) clearly demands the identity
$$
\tr q^2\=2\tr k+\frac6(\tr k)^2\Eqn\Matrixid
$$ 
to hold. This relation follows from
the selfduality relation (\SD) by contracting it with $F^{ijk}$.
Thus the on-shell value of ${*}\F$ simplifies to 
$$
({*}\F)^2=1+\frac{12}\tr k\punkt\Eqn\Fisfinal
$$
There are two more (related) consistency checks to be performed,
that concern chirality: the internal consistency of the selfduality 
relation (\SD) and the chirality projection implied by eqn. (\Proj).
By applying the selfduality relation twice, one obtains
a constraint on $({*}\F)^2$ which is identical to eqn. (\Fistwo).
The half-rank property of the matrix in eq. (\Proj) is also guaranteed
by selfduality and the actual value of $({*}\F)^2$.
This concludes the demonstration of $\k$-symmetry and the internal
consistency of the action, in the sense that it allows the consistent
restriction to a chiral self-interacting supersymmetric sector.
 
It should be stressed that neither eq. (\Matrixid) 
nor any other identities following from the selfduality relation may
be reinserted in the lagrangian. The field equations would then
change, so would the selfduality relation, and the entire intricate
web of consistency relations between equations of motion, selfduality
and $\k$-symmetry would break down.

Finally, we would like to show how our fields are related to the
ones in other formulations, where non-linear selfduality relations
also are encountered. We start by searching for a 3-form field $h$, formed
from $F$, which is linearly chiral. For simplicity, we chose positive
chirality, ${*}h\=h$, and set ${*}\F\=-N$ with $N$ positive.
An ansatz
\multi
$$
h_{ijk}=\varrho\bigl(F_{ijk}+\a q_i{}^lF_{jkl}\bigr)\equiv\varrho 
m_i{}^lF_{jkl}\komma\multieq\FieldRelationsa a 
$$
\intermulti
$$
F_{ijk}=\varrho^{-1}(m^{-1})_i{}^lh_{jkl}\komma\multieq\FieldRelationsb b
$$
gives $\a=-{1\/2N(N+1)}$. Note that eqn. (\FieldRelationsb)
is identical, up to normalisation, to the relation
between the linearly selfdual field strength and the closed one established
in ref. [\SixTensor], provided that $\varrho$ is a constant and that 
$m_{ij}=g_{ij}-{1\/2}h_i{}^{kl}h_{jkl}$.
Calculating the matrix $r_{ij}\equiv{1\/2}h_i{}^{kl}h_{jkl}$, one finds
using eqs. (\Matrixid) and (\Fisfinal) that
$$
r=\varrho^2q\bigl(1+\frac6\a^2\tr q^2+\frac3\a\tr k\bigr)
={2\varrho^2\/N(N+1)}q=-4\varrho^2\a q\punkt\eqn
$$
This provides a (highly non-trivial) check that the matrix $m$ has
the desired form for a constant value of $\varrho$, $\varrho\=\pm{1\/2}$,
reflecting only a difference in the normalisation of $F$ with respect to
the background tensor field.

\vskip2\parskip
\noindent\underbar{\it Acknowledgement:} We are grateful to Anders Westerberg
for discussions.

\refout 
\end